\documentclass[onecolumn]{pasj00}

\newcommand{\chobi}{\hspace{0.5mm}}

\newcommand{\NH}{{\mbox{\rm\chobi $N_H$}}}

\newcommand{\HI}{{\rm HI}}

\newcommand{\HALPHA}{{\rm H${\alpha }$}}

\newcommand{\UNITEFLUX}{{\rm ergs~cm$^{-2}$~s$^{-1}$}}
\newcommand{\UNITCPS}{{\rm counts~s$^{-1}$}}
\newcommand{\UNITPFLUX}{{\rm ph~cm$^{-2}$~s$^{-1}$}}

\newcommand{\UNITNH}{{\rm cm$^{-2}$}}

\newcommand{\UNITSOLARMASS}{{\it M$_{\odot}$}}

\newcommand{\KT}{{\it k$_{\rm B}T$}}

\newcommand{\SuzakuWebPage}{{http://www.astro.isas.jaxa.jp/suzaku/}}

\begin{document}
\SetRunningHead{Tsuru et al. }{X-ray Spectrum of 'the Cap' of M82}
\Received{2006/08/10}
\Accepted{2006/09/20}

\title{X-ray Spectral Study of the extended emission,
   'the Cap', located 11.6~kpc above the disk of M82}


\author{
   Takeshi Go \textsc{Tsuru}\altaffilmark{1},
   Midori \textsc{Ozawa}\altaffilmark{1},
   Yoshiaki \textsc{Hyodo}\altaffilmark{1},
   Hironori \textsc{Matsumoto}\altaffilmark{1},\\

   Katsuji \textsc{Koyama}\altaffilmark{1},
   Hisamitsu \textsc{Awaki}\altaffilmark{2},
   Ryuichi \textsc{Fujimoto}\altaffilmark{3},\\

   Richard \textsc{Griffiths}\altaffilmark{4},
   Caroline \textsc{Kilbourne}\altaffilmark{5},
   Kyoko \textsc{Matsushita}\altaffilmark{6},\\

   Kazuhisa \textsc{Mitsuda}\altaffilmark{3},
   Andrew \textsc{Ptak}\altaffilmark{7},
   Piero \textsc{Ranalli}\altaffilmark{8},
   and
   Noriko Y. \textsc{Yamasaki}\altaffilmark{3}
}
\altaffiltext{1}{
   Department of Physics, Kyoto University, Sakyo-ku, Kyoto 606-8502}
\email{tsuru@cr.scphys.kyoto-u.ac.jp, midori@cr.scphys.kyoto-u.ac.jp}
\altaffiltext{2}{
   Department of Physics,
   Ehime University,
   2-5 Bunkyo, Matsuyama, Ehime 790-8577}
\altaffiltext{3}{
   ISAS/JAXA, 2-1-1 Yoshinodai, Sagamihara 229-8510}
\altaffiltext{4}{
   Department of Physics, Carnegie Mellon University,
   Pittsburgh, PA 15213, USA}
\altaffiltext{5}{
   NASA Goddard Space Flight Center,
   Greenbelt, MD 20771, USA}
\altaffiltext{6}{
   Department of Physics, Tokyo University of Science, \\
   Kagurazaka, Shinjuku-ku, Tokyo 162-8601}
\altaffiltext{7}{
   Department of Physics and Astronomy, Johns Hopkins University, \\
   3400 N. Charles St., Baltimore, MD 21218, USA}
\altaffiltext{8}{
   Cosmic Radiation Lab., RIKEN, 2-1 Hirosawa, Wako, Saitama 351-0198}


%


\maketitle

\begin{abstract}
The extended X-ray emission from 'the Cap' region located
\timeform{11'} (11.6~kpc) above the disk of the starburst galaxy M82
has been observed with Suzaku and XMM-Newton. Owing to the good energy 
resolution and the large collecting area of the XIS on Suzaku, combined 
with similar properties of the 
EPIC instrument on XMM-Newton, we have clearly detected K-shell emission 
lines from O~VII, O~VIII, Ne~X, Mg~XI, Mg~XII and the Fe-L complex. 
Two optically-thin thermal plasma components are required to fit the 
observed X-ray spectra. We have determined the metal abundances of 
O, Ne, Mg, Si and Fe in this region for the first time. 
Their metal abundance ratios agree well with those of metal-poor stars 
and the model prediction of metals synthesized by type-II supernovae, 
but they are not consistent with the metallicities of
type-Ia supernovae. This result is support for the idea that the
origin of the metals in the Cap is type-II supernovae explosions
occurring in the starburst regions in the M82 galaxy. We discuss the 
possible contribution from sputtered dust grains to the metals in the Cap.
An emission line consistent with the C~VI transition of $n=4$ to $1$
at 0.459~keV is marginally detected, although it is not statistically
significant at the 99\% confidence level; the presence of this line 
would suggest charge-exchange processes in the Cap.
\end{abstract}

\section{Introduction} 
Starbursts inject a large amount of kinetic and
thermal energy into intergalactic space via outflows enriched 
with heavy metals synthesized in the massive stars that result in supernova
explosions. The metal-enriched ejecta
mix with the intergalactic medium and
now form part of the warm-hot intergalactic medium (WHIM) and
intra-cluster medium (ICM). Using ASCA results,
\citet{FukazawaPASJ98_Si_Fe_ICM_ASCA},
\citet{FinoguenovApJ00_ASCA_Cl_Metal} and
\citet{FinoguenovApJ01_ASCA_PoorCl_Metal} showed the importance of the
metal enrichment of the ICM by type-II supernova activity 
in starbursts at early epochs. The entropy excess in the ICM above the
self-similar prediction implies the importance of
non-gravitational heating. One of the plausible heat sources
is that from the superwinds which result from starbursts
\citep{PonmanNAT99_ICM_Entropy, LloydDaviesMN00_ICM_Entropy}. Thus,
starburst activity is a key phenomenon when trying to understand the metal
and thermal evolution of galaxies and clusters of galaxies. In spite
of the fact that we cannot see the initial starbursts directly, 
the observation of hot plasma currently being produced by the nearby 
starburst galaxies can reveal the basic processes and give us insight 
into the same processes at high redshift.

Since M82 is a prototypical starburst galaxy and one of the brightest
starburst galaxies in the local universe, it has been studied at
almost all wavelengths, ranging from low frequency radio up to TeV
$\gamma$-rays. Essentially all X-ray astronomical satellites have
observed M82. The extended
X-ray structure basically consists of
a multi-phase thermal plasma extending from the nuclear region toward
the halo, mainly along the minor axis of the galaxy
\citep{WatsonApJ84_M82_HRI, Fabbiano88ApJ_M82_N253_Ein,
 StricklandAA97_M82_ROS}. The point sources in M82 include the hyper
luminous X-ray source, M82 X-1, dominating X-ray luminosity above
$\sim 3$~keV \citep{Matsu99PASJ_M82AGN, Ptak99ApJL_M82AGN,
 Matsumoto2001_M82_CXO, KaaretMN01_M82_CXO_HRC}.

The spectral and spatial structure of the thermal plasma component of
M82 has been vigorously studied with CCDs or IGSPCs onboard ASCA,
BeppoSAX, Chandra and XMM-Newton \citep{Moran_Apj97_M82,
 Ptak1997_ASCA_M82, Tsuru1997_ASCA_M82, Cappi_AA99_SAX_M82_N253,
 WeaverApJ00_M82_N253_SBGII, GriffithsSci00_M82_ACIS,
 StevensMNL03_M82_EPIC, RanalliPH05_M82_XMM_GasAbundance}. The
results from the observation with RGS were reported by
\citet{ReadMN02_M82_RGS} and
\citet{OrigliaApJ04_M82_Star_Gas_Abundance}. The X-ray spectra are
well represented by a model consisting of multiple thermal plasma components
(\KT$\sim 0.3-1.5$~keV) with absorption due to the cool
matter in the M82 galaxy. The measurements of absolute metal abundances 
differ by a factor of several to ten among the measurements with ASCA,
XMM-Newton/EPIC and RGS \citep{Tsuru1997_ASCA_M82, ReadMN02_M82_RGS,
 StevensMNL03_M82_EPIC, OrigliaApJ04_M82_Star_Gas_Abundance}. The
abundances of O and Fe are lower than those of Si and S by a factor of
several, a common result from the ASCA and XMM-Newton/RGS
observations. Any combination of the metal production of type-Ia and
type-II supernovae cannot explain the abundance ratios among the
metals. Recently, \citet{RanalliPH05_M82_XMM_GasAbundance} have
reported that the metal abundances and their ratios significantly
change from the nucleus toward the halo. 
Additionally, the Fe abundance of $\sim 0.3$ solar was obtained
for the high temperature diffuse plasma in the core of M82
\citep{GriffithsSci00_M82_ACIS}.
Thus, the metal abundances of
the X-ray plasmas in the M82 galaxy itself are still under debate.

Searching for very extended X-ray emission far beyond the halo of
the M82 galaxy, \citet{Tsuru_PASJ90_GingaM82halo} have claimed
evidence for X-ray emission extending for 100~kpc around M82 from
scanning observations with the Ginga satellite. Extended
emission named 'the Cap' in the \HALPHA\ and soft X-ray bands 
was discovered at approximately $11'$, 
corresponding to 11.6~kpc at the distance of
3.63~Mpc, to the north of the center of M82
\citep{Devine1999_M82_NalphaCap, Lehnert1999_ROSAT_cap}.
Comparing the \HALPHA\ and the X-ray morphology, 
\citet{Lehnert1999_ROSAT_cap} have revealed a close relationship 
between the two emissions. This X-ray emission has been confirmed 
with XMM-Newton by \citet{StevensMNL03_M82_EPIC}. 
\citet{Lehnert1999_ROSAT_cap} explained its X-ray emission 
as the result of shock heating associated with an encounter
between the starburst-driven galactic superwind and a large
photoionized cloud in the halo of M82. 
\citet{Hoopes2005_GALEX_M82_N253} detected UV emission at the Cap
with GALEX and suggested that the most likely emission mechanism is
scattering of stellar continuum from the starburst in M82 by dust in the Cap.
The X-ray spectra of the Cap region obtained with ROSAT and XMM-Newton 
can be fitted with a thermal plasma model with a temperature of 
\KT$=0.80\pm 0.17$~keV or $0.65^{+0.69}_{-0.62}$~keV 
\citep{Lehnert1999_ROSAT_cap, StevensMNL03_M82_EPIC}. 
The metal abundance of this plasma is key
information in unraveling the origin and fate of the ejecta. 
However, no successful measurement has been made so far due to the limited
spectral resolution and/or statistics.

Thus, we made a 109~ks observation of the Cap with the Suzaku XIS. 
The XIS has good energy resolution, especially in the low energy band 
from 0.3 to1.0~keV, and low detector background. 
These properties have enabled us to obtain the
best quality X-ray spectrum in the Cap region and to elucidate 
the nature of the plasma, and especially 
to measure the metal abundances. We also analyzed the
archival data of the XMM-Newton observation of M82, 
in which the Cap was also observed in the same field of view. 
Throughout this paper, we adopt a
distance of $D=3.63$~Mpc to M82 \citep{Freedman1994_HST_M81_Dist}.
The solar metal abundances adopted in this paper are shown in
table~\ref{tab:SolarAbundance}.

\begin{table}
   \caption{The numbers of atoms per a H atom for the solar metal
     abundances adopted in this paper
     \citep{AndersGeCoA89_MeteoSolarAbundance}.
   }
   \label{tab:SolarAbundance}
   \begin{center}
     \begin{tabular}{cccccc}
       \hline\hline
       atom & number &atom & number & atom  & number \\
       \hline
       He & $0.0977$             &
       C  & $3.63\times 10^{-4}$ &
       N  & $1.12\times 10^{-4}$ \\
       O  & $8.51\times 10^{-4}$ &
       Ne & $1.23\times 10^{-4}$ &
       Mg & $3.80\times 10^{-5}$ \\
       Si & $3.55\times 10^{-5}$ &
       S  & $1.62\times 10^{-5}$ &
       Ar & $3.63\times 10^{-6}$ \\
       Ca & $2.99\times 10^{-6}$ &
       Fe & $4.68\times 10^{-5}$ &
       Ni & $1.78\times 10^{-6}$ \\
       \hline
     \end{tabular}
   \end{center}
\end{table}

\section{Observation and Data Reduction} 
\subsection{Data reduction of the Suzaku observation}
Three pointing observations aimed at the region including M82 and the
Cap were made with the X-ray CCD camera (XIS) onboard the Suzaku
satellite in October 2005 during the phase reserved for the Science Working
Group (the SWG phase). The details of the Suzaku satellite, 
the XRT and the XIS are found in \citet{Mitsuda2006_Suzaku}, 
\citet{Serlemitsos2006XRT} and \citet{Koyama2006XIS}, respectively. 
Data taken at an elevation angle
less than 5$^\circ$ from the Earth rim or during the passage through
the South Atlantic Anomaly were removed (Revision 0.7 data).

The reduction of the data from the 4 sensors of the XIS have been made
with XSELECT version 2.3 as follows. Firstly, we removed the telemetry
saturated data using GTI (good time interval) files. These GTI files
were provided by the XIS team through the Suzaku official web
page\footnote{\SuzakuWebPage}. Secondly we removed hot and flickering
pixels. Further, we did not use the data when the satellite attitude was not
stable after manoeuvre procedures. We confirmed from the light
curve that there were no data saturated periods during the
observation. After the filters and reductions were applied, we confirmed
that the data obtained from regions of blank sky showed no significant time
variability.

We compiled the data from the 3x3 and 5x5 modes of the three
observations into one. After these filters were applied, a net
exposure time of 109~ks was left for both BI and FI. The observation
logs are given in table~\ref{tab:ObsLog}.


\subsection{Data reduction of the XMM-Newton observation}
M82 and the Cap were observed twice with XMM-Newton. 
The details of the XMM-Newton satellite, the EPIC-MOS and 
pn instruments can be found in \citet{Jansen2001_XMM}, 
\citet{Turner2001_EPIC_MOS} and \citet{Struder2001_EPIC_PN}, respectively. 
The effective exposure after the screening of the first observation 
on 2001 May 6 was 20-30~ks \citep{ReadMN02_M82_RGS, 
 OrigliaApJ04_M82_Star_Gas_Abundance, StevensMNL03_M82_EPIC}. 
Out of the total exposure time of 100~ks in
the second observation on 2004 April 21, an effective exposure of
59-65~ks was left, which is two or three times longer than that of the
first one. The energy resolution of the EPIC-MOS was significantly
improved after 2002 November by changing the detector temperature of
EPIC from $-100{\rm C^\circ}$ to $-120{\rm C^\circ}$
\citep{KirschSPIE05_XMM_pn_MOS}.  For these reasons, we concentrate on
the second observation in this paper.

The data from the EPIC-MOS and pn instruments have been processed with
the standard procedures of the XMM-Newton SAS (Science Analysis System
6.0.0). Periods of high background were filtered out, i.e.
EPIC-MOS data with a counting rate above $0.3$~\UNITCPS\ 
in the energy band 10-15~keV were removed. 
As for the EPIC-pn, we did not use data
with a counting rate above $1.0$~\UNITCPS\ in the same energy band. 
After these filters were applied, net exposures of
59~ks and 65~ks were left for the EPIC-MOS and pn instruments,
respectively. The observation logs are also given in
table~\ref{tab:ObsLog}.

\begin{table}
   \caption{The logs of Suzaku and XMM-Newton observations of M82.}
   \label{tab:ObsLog}
   \begin{center}
     \begin{tabular}{lcccc}
       \hline\hline
       Instrument & Seq. No. & Observation & Effective \\
                  &          & Start Date  & Exposure  \\
       \hline
       XIS        &100033010  & 2005/10/04 &  36.8 ks \\
                  &100033020  & 2005/10/19 &  41.7 ks \\
                  &100033030  & 2005/10/27 &  30.5 ks \\
       EPIC-pn    &0206080101 & 2004/04/21 &  53.0 ks \\
       EPIC-MOS   &0206080101 & 2004/04/21 &  65.0 ks \\
       \hline
     \end{tabular}
   \end{center}
\end{table}

\section{Analysis and Results} 
\subsection{Imaging Analysis and Overall Structure}
\begin{figure}
   \begin{center}
     \FigureFile(100mm,80mm){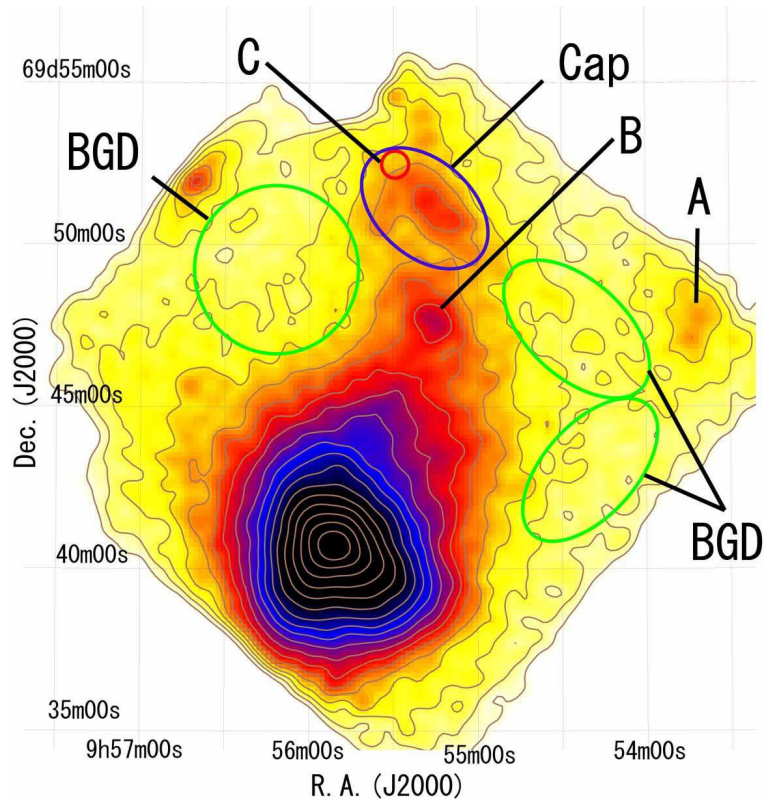}
   \end{center}
   \caption{X-ray image obtained with the Suzaku XIS1 (BI) in the energy
     band  0.3-2~keV. The attitude correction  has been already
     done by comparison with the XMM-Newton/EPIC-MOS image. 
     The image has been smoothed with $\sigma=3$ pixels. 
     The regions used for the spectra of the Cap and  BGD are shown. 
     The position of the sources A, B and C are also indicated.
   }
   \label{fig:Suzaku_BI_Image}
\end{figure}

\begin{figure}
   \begin{center}
     \FigureFile(100mm,80mm){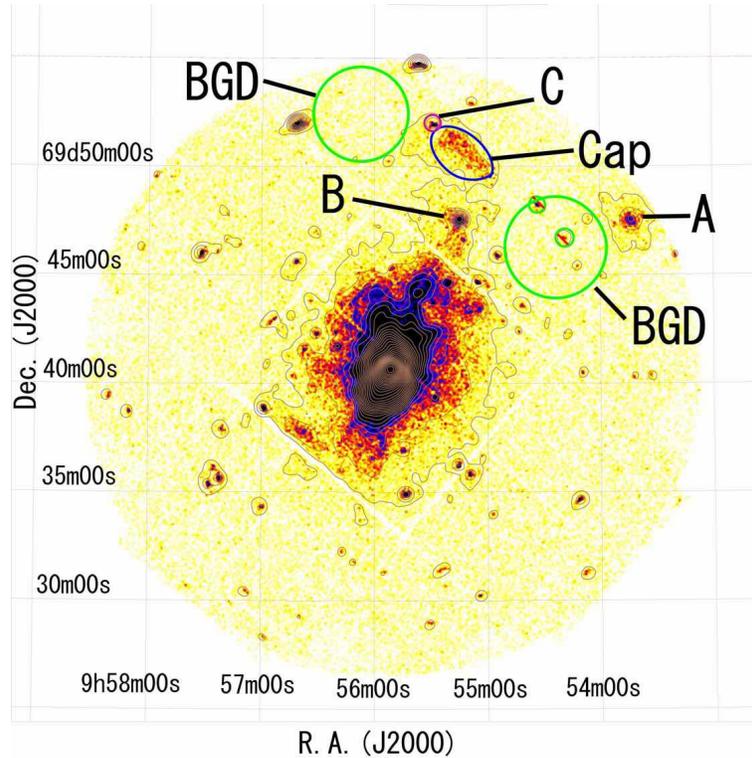}
   \end{center}
   \caption{X-ray image obtained with EPIC-MOS1 and MOS2 in the energy
     band 0.3-2~keV. The image is smoothed with $\sigma=3$ pixels.
     The regions used for the spectra of the Cap and BGD are
     shown. The position of the sources A, B and C are also indicated.
     The data from the two small circles inside one of the BGD
     regions were not used because point sources are located there.
   }
   \label{fig:XMM_MOS_Image}
\end{figure}

Figures~\ref{fig:Suzaku_BI_Image} and \ref{fig:XMM_MOS_Image} show
X-ray images of M82 in the energy band  0.3-2~keV obtained with
Suzaku XIS1 (BI) and XMM-Newton MOS1$+$MOS2, respectively. The X-ray
emission of the M82 galaxy in the energy band  3-10~keV is dominated  
by
the hyper-luminous X-ray source M82 X-1 \citep{Matsu99PASJ_M82AGN,
   Matsumoto2001_M82_CXO}. The nominal positions of the X-ray peak
(henceforth 'M82 X-1') in the images in the energy band 3-10~keV obtained
with Suzaku XIS and XMM-Newton EPIC-MOS are $(\alpha_{2000},
\delta_{2000}) = (9^{\rm h}55^{\rm m}47.0^{\rm s}, 69^\circ 40' 07'')$
and $(9^{\rm h}55^{\rm m}50.6^{\rm s}, 69^\circ 40' 45'')$,
respectively. Another bright point source 'B' seen in the XIS image at
$(9^{\rm h}55^{\rm m}11.5^{\rm s}, 69^\circ 46' 55'')$ in the nominal
Suzaku coordinates can be identified with a point source at $(9^{\rm
   h}55^{\rm m}14.8^{\rm s}, 69^\circ 47'36'')$ in the EPIC-MOS
image. The absolute pointing accuracy of XMM-Newton is better than
$1''$. Therefore, the Suzaku nominal coordinate system is systematically
shifted by $(3.7^{\rm s}, 39.5'')$ from the XMM-Newton coordinates.
This spatial offset is within the calibration accuracy 
at the time of writing \citep{Serlemitsos2006XRT}. 
Thus, we tuned the Suzaku coordinates by shifting $(-3.7^{\rm s}, -39.5'')$.

Clear diffuse X-ray emission is seen in the Cap region of both
figures~\ref{fig:Suzaku_BI_Image} and \ref{fig:XMM_MOS_Image},
and this is the prime target of this paper. In the XMM-Newton image, 
a point source denoted as 'C' is detected at the position 
$(\alpha_{2000},\delta_{2000})=(09^{\rm h}55^{\rm m}27.7^{\rm s}, 
+69^\circ 52' 1'')$, is consistent with the point source 
seen at $(09^{\rm h}55^{\rm m}29.6^{\rm s}, +69^\circ 51' 55'')$ 
in the ROSAT observation \citep{Lehnert1999_ROSAT_cap}. 
This point source is identified with a stellar object 
having an R magnitude of 12.2 \citep{Lehnert1999_ROSAT_cap}. 
The source C is unclear in the Suzaku image of 
figure~\ref{fig:Suzaku_BI_Image}. As described in the following section, 
point source C has the low flux of $1.6\times 10^{-14}$~\UNITEFLUX\ 
in the energy band 0.3-3~keV, 
and is therefore too faint to be detected with Suzaku.

In addition to the X-ray emission at the Cap, diffuse X-ray emission is
clearly seen in the connecting region between the Cap and the M82
galaxy (figures~\ref{fig:Suzaku_BI_Image} and \ref{fig:XMM_MOS_Image})
as already reported by \citet{StevensMNL03_M82_EPIC}. Our preliminary
analysis of the XMM-Newton data show that this diffuse emission has a
flux of $8.9\times 10^{-14}$~\UNITEFLUX\ in the energy band   
0.3-3~keV.

We note that the source 'A' at $(\alpha_{2000},\delta_{2000}) 
= (09^{\rm h}53^{\rm m}43.5^{\rm s},+69^\circ 47' 35'')$ is also 
a diffuse X-ray source. \citet{GalAJ03_Opt_Northern_Cl_Catalog} reported 
that a cluster of galaxies, NSC~J095337$+$694751 at a red-shift of
$z=0.211$, is located at the position $(\alpha_{2000}, \delta_{2000}) 
= (09^{\rm h}53^{\rm m}37.53^{\rm s}, +69^\circ 47' 50.9'')$. 
Source 'A' could therefore be identified with the cluster 
in spite of the separation of $\sim 0.6'$ between the two positions.

\subsection{Spectrum Analysis of EPIC data of the Cap}
\label{SubSec:SpecAnaCapEPIC}
We analyzed the EPIC-MOS and pn data for the Cap and a background area 
near the Cap, inside the bounding regions shown in 
figure~\ref{fig:XMM_MOS_Image}. 
Before background subtraction, the total counts from the Cap
region in the energy band  0.3-3~keV were 2496 and 3610 counts for
the EPIC-MOS and pn, respectively. After subtraction of the background
data, the flux from the Cap region was $1.2\times
10^{-13}$~\UNITEFLUX\ in the energy band  0.3-3~keV. The background
subtracted spectra are given in figure~\ref{fig:XMMSpec} along with the best
fitted model spectra. The details of the spectral fits will be
described in section~\ref{SubSect:SimpleBGDSubt}.

\subsection{Spectral Analysis of the EPIC data on Point Source C}
\label{SubSec:SpecAnaPointSourceC}
The primary purpose of this paper is to obtain a high quality 
X-ray spectrum of the Cap. Because it is difficult to resolve point
source C from the diffuse component of the Cap using the Suzaku XIS, we
analyzed the EPIC data to obtain the spectrum of point source C.
From the small region including point source C shown in
figure~\ref{fig:XMM_MOS_Image}, 268 and 402 counts were obtained for
the EPIC-MOS and pn, respectively. No significant time variability was
detected within the XMM-Newton exposure. After subtraction of
the background data from the same region as that used for the Cap, the
flux of source C was $1.6\times 10^{-14}$~\UNITEFLUX\ in the
same band. Thus, the flux of source C is 13\% of that of the Cap, 
indicating that the contribution from this source
is relatively minor in the whole Cap region.

We fitted the EPIC-MOS and pn spectra of point source C in the
energy band 0.3-3~keV and obtained an acceptable result with an
empirical model consisting of two thin thermal plasma components, in
which we adopted the vMEKAL model incorporated in XSPEC version 11.3
\citep{Mewe1985_MEKAL}. The best fit parameters are given in
table~\ref{tab:SimpleFitResults}. 
Its thermal nature of the spectrum is consistent with a late-type dwarf star, 
which is suggested by \citet{Lehnert1999_ROSAT_cap}. 
This model spectrum describing 
point source C will be included in every spectral fit for the Suzaku
spectra of the Cap in the following analysis.

\subsection{Procedure used for the Spectral Analysis of the XIS data}
\label{SubSec:ProcSpecAnaXIS}
Previous observations of the M82 galaxy showed that its X-ray spectrum is
well described by a thermal plasma model, in which prominent
thermal emission lines of several elements are detected
\citep{Tsuru1997_ASCA_M82, Ptak1997_ASCA_M82, ReadMN02_M82_RGS,
   StevensMNL03_M82_EPIC}. Thus, in order to check the XIS gain using
these lines, we produced an X-ray spectrum for each XIS sensor by
collecting counts within a circular region with a radius of 2.48~arcmin
around M82. We used the versions of February 13, 2006 and
May 28, 2006 for the response (RMF) and xissimarfgen (ARF)
\citep{Ishisaki2006ARF}, respectively. We detected prominent K
emission lines of O~VII, O~VIII, Ne~X, Mg~XI and Mg~XII in the
spectra. We fitted these spectra with gaussian functions and a power law
model empirically describing the emission lines and the continuum
component. Comparing the center energies of the lines with the
theoretical values, we found that the discrepancies are less than 4~eV, which
indicates that the absolute energy scale of the XIS data is well
calibrated.

The optical blocking filters (OBF hereafter) of the XIS sensors are
unfortunately contaminated by outgas from the satellite
\citep{Koyama2006XIS}. The results from the XIS calibration show
that the column density of the contaminant can be described as a function
of the radial distance from the center of the XIS field of view (FOV). 
The vignetting of the
XRT is also given as a function of radius from the FOV center
\citep{Serlemitsos2006XRT}. Thus, we selected the Cap and the BGD
regions to have the same distance from the FOV center 
in order to reduce the systematic errors due to the
calibration uncertainties from the contaminant and the vignetting
function (figure~\ref{fig:Suzaku_BI_Image}). The ARFs are calculated
with xissimarfgen on the assumption that the BGD and the Cap are
uniformly diffuse and point like, respectively.

We performed background subtraction on the XIS spectra of the Cap with
the two methods described below. The first one is a simple method in
which we collected background spectra from the BGD region shown in
figure~\ref{fig:Suzaku_BI_Image} and subtracted them from the ones of
the Cap (method 1). The background subtracted XIS FI and BI spectra of 
the Cap are shown in figure~\ref{fig:SuzakuSimpleBGDsubtSpec}, 
where the FI data from the three instruments (XIS0, XIS2 and XIS3) are
combined. Note that the spectra also include the contribution
from point source C detected with XMM-Newton. Subtraction of
a model spectrum for this source will be done 
during the spectral fitting for the Cap.

In the second method, we fitted spectra of the X-ray background and
included this model in the fits to the X-ray spectra of the Cap
as mentioned below (method 2). The model describing point source C is also
included in the fitting. In the first step of this method, we
made spectra of the night earth data having the same detector coordinates 
as the Cap region and used these as the detector background.
We adopted night earth data collected by the
Suzaku XIS team from September 2005 to May 2006 and screened in the
Revision 0.7 procedure. After the filters were applied, a net
exposure time of 797~ks was  left for both FI and BI detectors. We subtracted
the spectra of the night earth (as the detector background) from those
of the Cap. We followed the same procedure for the background region.
Hereafter we call the night-earth-subtracted spectra of the Cap and
background region 'the CAP-NTE' and 'the BGD-NTE', respectively.

The BGD-NTE is thought to consist of the cosmic X-ray background (CXB)
and the soft X-ray background (SXB). To determine these components, we
fitted the BGD-NTE spectra. The CAP-NTE spectra comprise the X-ray
emission from the Cap, point source C, the CXB and the SXB, the
last two of which can be determined with the BGD-NTE. Thus, we made
spectral fits to the BGD-NTE using the model spectrum describing the
CXB plus SXB. The model spectrum of point source C was already
obtained in section~\ref{SubSec:SpecAnaPointSourceC}. Including the
models for the BGD-NTE and point source C in the fits for the
CAP-NTE, we finally obtained the X-ray spectra of the Cap.

Then, if we were to adopt the best fit model for the BGD-NTE as the
'X-ray background' and use it in the spectral fit to the CAP-NTE, it
would result in ignoring the statistical errors included in the
observed spectra of BGD-NTE. Thus, we conducted simultaneous fits
to the spectra of the CAP-NTE and the BGD-NTE linking the model
parameters describing the X-ray background (the CXB and SXB) to be
common in both of the fits. This method allows us to include
the statistical errors of the CAP-NTE into the fit to the spectrum
of the Cap as a contribution to the errors of the model parameters. 
In addition to
the fits to the XIS spectra in the second method, we included the EPIC
spectra of the Cap and made joint spectral fits in order to
improve the overall fit. The results are given in the next section.

\subsection{Results from the Spectral Analyses of the Cap}
\subsubsection{Using Simple Background Subtraction (method 1)}
\label{SubSect:SimpleBGDSubt}
\begin{figure}
   \begin{center}
     \FigureFile(120mm,80mm){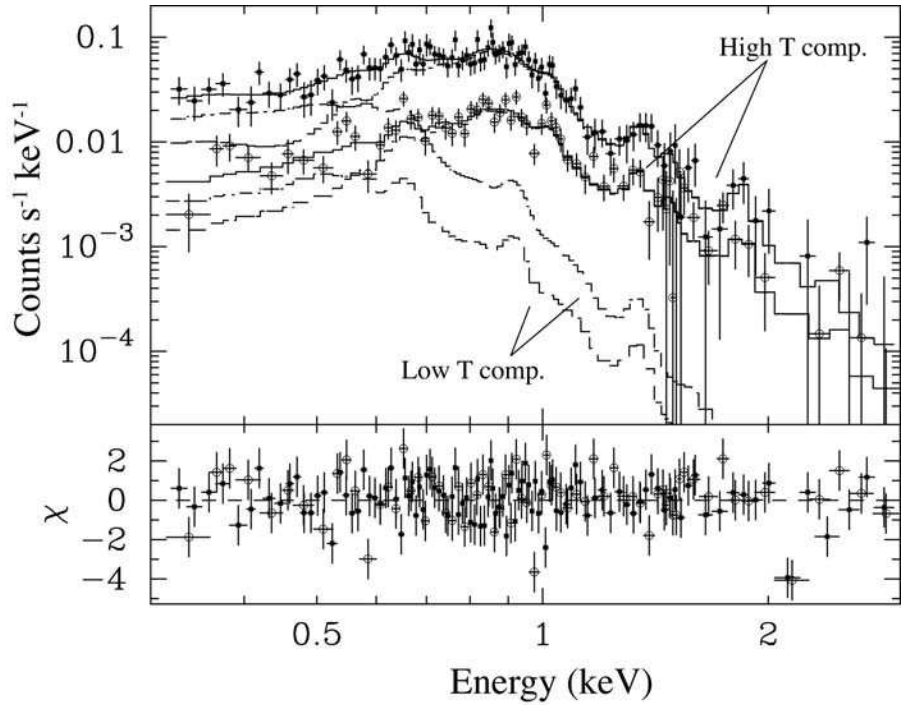}
   \end{center}
   \caption{The simple background-subtracted EPIC spectra of the Cap
     are shown with the best fit 2vMEKAL model spectra. Black squares
     indicate the EPIC-pn spectrum. The open circles show the combined
     spectrum of the two MOS (MOS1 and MOS2) with the normalization
     for one MOS sensor.
   }
   \label{fig:XMMSpec}
\end{figure}

\begin{figure}
   \begin{center}
     \FigureFile(120mm,80mm){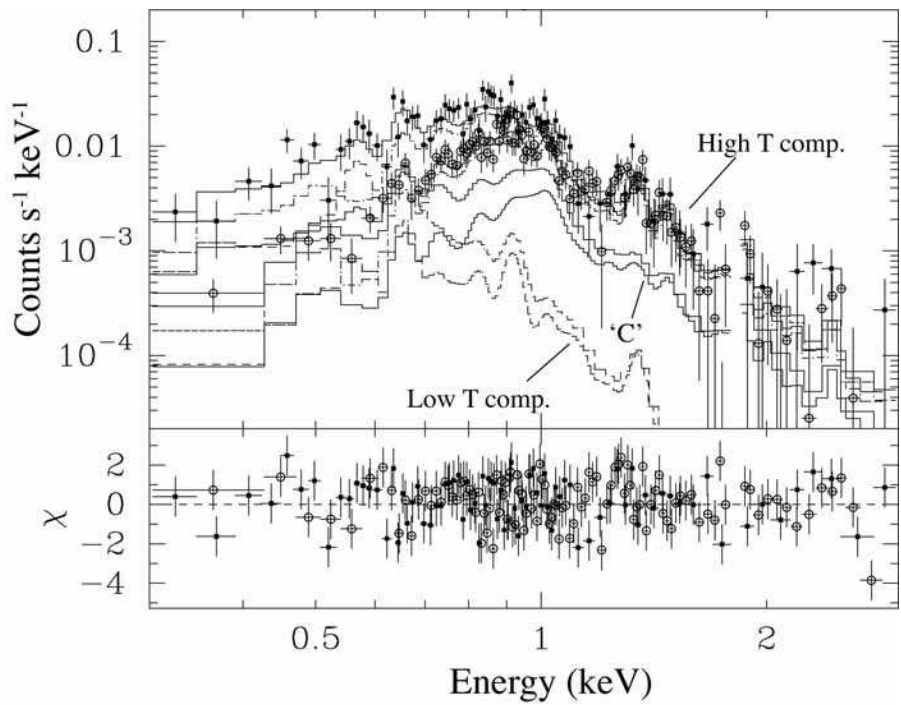}
   \end{center}
   \caption{The simple background-subtracted XIS spectra of the Cap are
     shown with the best fit 2vMEKAL model spectra. Black squares
     indicate the BI (XIS1) spectrum. The open circles show the
     combined spectrum of the three FI (XIS0,2,3) with the normalization
     for one FI sensor.
   }
   \label{fig:SuzakuSimpleBGDsubtSpec}
\end{figure}

Figures~\ref{fig:XMMSpec} and \ref{fig:SuzakuSimpleBGDsubtSpec} show
the spectra of the Cap obtained with the EPIC and XIS with the best fit
model spectra. In the XIS spectra of the Cap, several prominent K
emission lines are seen and identified 
with O~VII (0.57~keV), O~VIII (0.65~keV),
Ne~X (1.02~keV), Mg~XI (1.34~keV), Mg~XII (1.47~keV) and the Fe-L
complex. \citet{StevensMNL03_M82_EPIC} reported that the Cap
spectra obtained with the first observation of XMM-Newton can be
represented by a single thin thermal plasma model with the metal
abundances fixed at the solar values. Therefore, we started
fitting the XIS spectra with a single absorbed thin thermal plasma
model with a temperature of \KT$\sim 0.6$~keV plus the best-fit
model for point source C. We used
vMEKAL in XSPEC for the thermal plasma model \citep{Mewe1985_MEKAL}.
The energy band above 3~keV was ignored, since no significant flux was
detected there. With this fit, we did not reach reasonable results even
when the metal abundances were left free, since a positive residual was
found at O~VII (1vMEKAL in table~\ref{tab:SimpleFitResults}). 
We also found a positive residual at the O~VII K emission line energy 
(0.57~keV) from the fit to the EPIC spectra with the same model, 
although the overall structure of the EPIC spectra is 
described well by a single thermal plasma model.
The line emissivity of O~VII at a temperature of \KT$\sim 0.6$~keV
is only $\sim 2$\% of that at \KT$\sim 0.17$~keV, which is the
temperature giving the peak emissivity of the O~VII line. Thus, this
result suggests the need for an additional component to explain the
O~VII line.

Firstly, we fitted a thermal plasma model plus a
narrow gaussian line to match the O~VII line (1vMEKAL+1GAUS). 
We found that this gave a reasonable result, as 
shown in table~\ref{tab:SimpleFitResults}.
The center energy of the gaussian line is consistent with the
theoretical value of O~VII K emission at $0.57$~keV. 
$\chi^2/{\rm d.o.f.}$ is reduced from $503.2/361$ for the 1vMEKAL
model to $479.4/359$ for 1vMEKAL$+$1GAUS, which allows us to 
conclude the introduction of 1GAUS is significant from the statistical 
point of view. 
However, the upper limit on the absorption column obtained from this fit
contradicts the results from \HI\ observations, 
as shown by the following argument. The
column density of the cool matter in our galaxy toward the M82 region
is \NH$=4.0\times 10^{20}$~\UNITNH\ \citep{Dickey1990_NH}. The M81/M82
group are shown to be embedded in a large complex cloud of neutral H
\citep{Cottrel1977_HI, Appleton1981_HI}. It has been shown that there is an
apparent hole in the \HI\ emission and thus the Cap region does
not show any associated \HI\ emission \citep{Lehnert1999_ROSAT_cap,
 StevensMNL03_M82_EPIC}. In fact, the column density of the \HI\ cloud 
in the Cap region is less than $2.7\times
10^{19}$~\UNITNH\ \citep{Yun1993_M82_HI, Yun1994_M81group_HI}.

We have to take into account the contaminant on the XIS OBFs 
and its associated uncertainty. 
The column densities of the carbon contaminant on the OBFs at
the XIS detector positions corresponding with the Cap observations is 
estimated to be $7.08\times 10^{17}$, $8.01\times 10^{17}$, 
$1.08\times 10^{18}$ and $2.15\times 10^{18}$~\UNITNH\ 
for XIS0, XIS1, XIS2, and XIS3, respectively. The
column densities of O are assumed to be $1/6$  those of C. The
thickness of the contaminant on the XIS OBFs still has
some uncertainty, estimated to be $\pm 20\%$ 
at the time of writing \citep{Koyama2006XIS}. 
In the energy band of $0.3$-$1$~keV, the absorption due to the
contaminant corresponds to the one due to \HI\ with
\NH$=2.5\times 10^{20}\sim 7.6\times 10^{20}$~\UNITNH. Thus, the
uncertainty in the contaminant thickness is equivalent to a galactic
absorption of $\Delta$\NH$=(0.5\sim 1.5)\times 10^{20}$~\UNITNH. Note
that the uncertainty equivalent to the galactic absorption for XIS1
(BI) is $\Delta$\NH$=0.5\times 10^{20}$~\UNITNH; this detector is the most
sensitive to the absorption column. Even if
the uncertainty on the contaminant is taken into account, the upper
limit on the absorption column density derived from the spectral
fit is still lower than that of the galactic \HI. Thus, we
conclude that the model of a thermal plasma plus a narrow gaussian
line is physically inadequate for the description of the observed
Cap spectra.

The above result is also evidence for the existence of an additional
continuum component in the low-energy X-ray band. So we next tried a
model consisting of two thin thermal plasma components (2vMEKAL)
having common metal abundances, in which the lower temperature
component explains the additional continuum component and the O~VII line. 
The fit gave a satisfactory result, 
in which $\chi^2/{\rm d.o.f.}$ of 2vMEKAL is 
nearly equal to that of 1vMEKAL$+$1GAUS (table~\ref{tab:SimpleFitResults}). 
The upper limit on the column density of the absorbing cool matter is 
consistent with the sum of the column densities of the cool matter 
in our galaxy toward the M82 region (\NH$=4.0\times 10^{20}$~\UNITNH) 
and the upper limit on the column density of the \HI\ cloud 
in the Cap region ($<2.7\times 10^{19}$~\UNITNH). 
In the following, we fix the whole absorption
column density to be \NH$=4.0\times 10^{20}$~\UNITNH\ in the fits.

\begin{longtable}{lcccc}
   \caption{The results from the simple background-subtracted spectra
     are described (method 1 in the text). The component normalization
     of the vMEKAL model is $\left(10^{-14}/4\pi D^2\right)\int{n_{\rm
         H}n_{\rm e}}dV{\rm cm^{-5}}$, where $V$ is the volume and $D$ the
     distance. The metal abundances are given in units of the solar
     value. The abundances of the metals except the ones given 
     in the table are fixed to be solar. The unit of absorption 
     column is $10^{20}$~\UNITNH. The
     normalization of the gaussian line is defined as an unabsorbed
     photon flux in units of $10^{-6}~{\rm ph\ cm^{-2}\ s^{-1}}$.
     
   }\label{tab:SimpleFitResults}
    \hline
    \endhead
    \hline
    \endfoot
    Data              & point source 'C'       & \multicolumn{3}{c}{The Cap}                             \\
    \hline
    Instruments       & EPIC                   & \multicolumn{3}{c}{EPIC \& XIS}                         \\
    Model             & 2vMEKAL                & 1vMEKAL                 & 1vMEKAL+1GAUS             & 2vMEKAL                \\
    \hline
    \NH\              & $0$ (fixed)            & $<1.43$ (90\%)          & $<1.26$ (90\%)            & $<6.4$ (90\%) \\
                      &                        & $<2.54$ (99\%)          & $<1.91$ (99\%)            & $<9.0$ (99\%) \\
    \KT\ (keV)        & $0.28^{+0.05}_{-0.05}$ & $0.63^{+0.02}_{-0.02}$  & $0.63^{+0.02}_{-0.01}$    & $0.21^{+0.06}_{-0.09}$ \\
    Norm.             & $2.66\times 10^{-6}$   & $5.88\times 10^{-5}$    & $5.27\times 10^{-5}$      & $1.01\times 10^{-5}$   \\
    \KT\ (keV)        & $0.99^{+0.11}_{-0.16}$ & ...                     & ...                       & $0.63^{+0.02}_{-0.02}$ \\
    Norm.             & $4.48\times 10^{-6}$   & ...                     & ...                       & $4.51\times 10^{-5}$   \\
    O                 & $1$ (fixed)            & $1.18^{+0.27}_{-0.26}$  & $1.33^{+0.11}_{-0.27}$    & $1.15^{+1.00}_{-0.27}$ \\
    Ne                & $1$ (fixed)            & $0.84^{+0.31}_{-0.26}$  & $0.99^{+0.28}_{-0.21}$    & $1.31^{+0.43}_{-0.37}$ \\
    Mg                & $1$ (fixed)            & $1.17^{+0.31}_{-0.29}$  & $1.37^{+0.28}_{-0.23}$    & $1.81^{+1.02}_{-0.41}$ \\
    Si                & $1$ (fixed)            & $0.60^{+0.40}_{-0.36}$  & $0.71^{+0.40}_{-0.38}$    & $0.96^{+0.51}_{-0.48}$ \\
    Fe                & $1$ (fixed)            & $0.36^{+0.07}_{-0.06}$  & $0.42^{+0.05}_{-0.03}$    & $0.54^{+0.13}_{-0.07}$ \\
    LineE (keV)       & ...                    & ...                     & $0.575^{+0.010}_{-0.016}$ & ...                    \\
    Norm.             & ...                    & ...                     & $3.86^{+1.43}_{-1.16}$    & ...                    \\
    \hline
    $\chi^2$/d.o.f.   & $22.55/22$             & $503.2/361$             & $479.4/359$               & $478.2/359$            \\
    \hline
\end{longtable}

\subsubsection{Simultaneous Spectral Fitting of the Cap and the Blank Sky 
  (method 2)}
\label{Sect:SimFitCapBlank}
We now describe the results from the second method, employed with the
purpose of improving the fit, in which simultaneous fits
to the spectra of the CAP-NTE and the BGD-NTE were made.

Firstly, we fitted the BGD-NTE spectra. Specifically, we adopted an
empirical X-ray background model consisting of two thin thermal plasma
components and a power law component (2vMEKAL$+$POW). The former
thermal components represent the soft X-ray background mainly due to
the hot plasmas in our galaxy whereas the latter power law model fits
the cosmic X-ray background (CXB) of extragalactic origin. The whole
absorption was fixed at a column density of \NH$=4.0\times
10^{20}$~\UNITNH, which was described in
section~\ref{SubSect:SimpleBGDSubt}. We found that the model with the
metal abundances fixed to solar values fails to represent the BGD-NTE
spectra in the energy band of 0.3-7~keV. We therefore let the
abundances of Ne, Mg, and Fe vary while the metal abundances were
fixed to be common between the two thin thermal plasmas. This model
gave a reasonable fit (table~\ref{tab:SimultFit}). The normalization
and the photon index of the power law component are consistent with
the values previously reported by \citet{Kushino2002_ASCA_CXB_SXB}
and \citet{Lumb2002_XMM_CXB_SXB}.

Following the procedure described in section~\ref{SubSec:ProcSpecAnaXIS},
we performed simultaneous fits to the XIS spectra of the CAP-NTE and
the BGD-NTE. We also included the EPIC spectra
obtained in section~\ref{SubSect:SimpleBGDSubt} in the fits. Since
the X-ray spectrum of the Cap is soft, we used the data in the energy
band  0.3-3~keV. The photon index of the CXB power law component
was fixed to the best-fit values obtained above
with the normalization left free, since we use the data only below
3~keV, where the contribution from the CXB is minor. The model
spectrum describing point source C was fixed as before.

We first tried the model comprising a thermal plasma component and a narrow
gaussian line explaining the O~VII emission line, leaving the
absorption column free (1vMEKAL+1GAUS). The results are given in
table~\ref{tab:SimultFit}. We found the upper limit on the absorption
column to be lower than that of our galaxy again, which confirms the
conclusion obtained in section~\ref{SubSect:SimpleBGDSubt}.

We next conducted the spectral fitting with two thermal plasma
components (2vMEKAL). The whole absorption was fixed at a column
density of \NH$=4.0\times 10^{20}$~\UNITNH\ based on the results shown
in section~\ref{SubSect:SimpleBGDSubt}. The fit gave satisfactory
results with nearly equal $\chi^2/{\rm d.o.f.}$ 
to that of 1vMEKAL$+$1GAUS, consistent with the one
obtained in section~\ref{SubSect:SimpleBGDSubt} while the errors of
the parameters became smaller (table~\ref{tab:SimultFit}).

\begin{longtable}{lccccc}
   \caption{The results from the simultaneous spectral fits to the
     Cap and the blank sky are described (method 2 in the text).
     Normalization for the fitting of 'BGD-NTE' is given for
     the actual BGD region shown in figure~\ref{fig:Suzaku_BI_Image},
     where the ratio of BGD area to that of the Cap
     region is $4.34$. The model component normalizations are:
     for the vMEKAL and vAPEC models
     $\left(10^{-14}/4\pi D^2\right)\int{n_{\rm H}n_{\rm e}}dV{\rm
       cm^{-5}}$, where $V$ is the volume and $D$ the distance, while for
     the power-law model the unit of normalization is ${\rm
       keV^{-1}cm^{-2}s^{-1}}$ at $1{\rm keV}$. The metal abundances
     are given in solar units. The abundances of the metals 
     except the ones given in the table are fixed to be solar.
     The unit of absorption column is $10^{20}$~\UNITNH. 
     The normalization of the gaussian line is defined as an unabsorbed 
     photon flux with a normalization unit of $10^{-6}$~\UNITPFLUX.
   }
   \label{tab:SimultFit}

    \hline
    \endhead
    \hline
    \endfoot
    Data               & BGD-NTE                  &  \multicolumn{4}{c}{The Cap} \\
    \hline
    Model              & 2vMEKAL                  & 1vMEKAL                     & 2vMEKAL
                                                  & 2vAPEC                      & 3vMEKAL                   \\
                       & +1POW                    & +1GAUS                      &
                                                  &                             &                           \\
    \hline
    \NH                & $4$ (fixed)              & $<1.28$ (90\%)             & $4$ (fixed)
                                                  & $4$ (fixed)                & $4$ (fixed)               \\
                       &                          & $<1.89$ (99\%)             & 
                                                  &                            &                           \\
    \KT (keV)          & $0.64^{+0.04}_{-0.04}$   & $0.63^{+0.01}_{-0.01}$     & $0.63^{+0.02}_{-0.02}$
                                                  & $0.64^{+0.02}_{-0.02}$     & $0.71^{+0.04}_{-0.08}$    \\
    Norm.              & $3.57\times 10^{-5}$     & $6.18\times 10^{-5}$       & $5.96\times 10^{-5} $
                                                  & $6.22\times 10^{-5}$       & $3.61\times 10^{-5}$      \\
    \KT (keV)          & $0.20^{+0.02}_{-0.02}$   & ...                        & $0.20^{+0.01}_{-0.01}$
                                                  & $0.23^{+0.03}_{-0.05}$     & $0.13^{+0.08}_{-0.03}$    \\
    Norm.              & $3.85\times 10^{-5}$     & ...                        & $1.45\times 10^{-5}$
                                                  & $2.48\times 10^{-5}$       & $1.56\times 10^{-5}$      \\
    \KT (keV)/$\Gamma$ & $1.27^{+0.09}_{-0.09}$   &  ...                       & ...
                                                  & ...                        & $0.44^{+0.10}_{-0.17}$    \\
    Norm.              & $4.26\times 10^{-5}$     &  ...                       & ...
                                                  & ...                        & $2.40\times 10^{-5}$      \\
    O                  & $1$ (fixed)              & $1.30^{+0.13}_{-0.18}$     & $1.11^{+0.33}_{-0.32}$
                                                  & $0.76^{+0.27}_{-0.20}$     & $1.36^{+0.29}_{-0.47}$    \\
    Ne                 & $1.29^{+0.87}_{-0.68}$   & $0.97^{+0.21}_{-0.19}$     & $1.10^{+0.54}_{-0.40}$
                                                  & $1.36^{+0.36}_{-0.40}$     & $1.06^{+0.44}_{-0.44}$    \\
    Mg                 & $2.17^{+1.28}_{-0.77}$   & $1.43^{+0.24}_{-0.22}$     & $1.59^{+0.67}_{-0.48}$
                                                  & $1.37^{+0.51}_{-0.36}$     & $1.79^{+0.72}_{-0.56}$    \\
    Si                 & $1$ (fixed)              & $0.92^{+0.37}_{-0.36}$     & $1.03^{+0.58}_{-0.45}$
                                                  & $0.86^{+0.43}_{-0.36}$     & $1.15^{+0.61}_{-0.49}$    \\
    Fe                 & $0.51^{+0.24}_{-0.13}$   & $0.42^{+0.03}_{-0.02}$     & $0.49^{+0.07}_{-0.10}$
                                                  & $0.44^{+0.08}_{-0.10}$     & $0.54^{+0.08}_{-0.14}$    \\
    LineE (keV)        & ...                      & $0.570^{+0.003}_{-0.014}$  & ...
                                                  & ...                        & ...                       \\
    Norm.              & ...                      & $5.87^{+1.46}_{-1.63}$     & ...
                                                  & ...                        & ...                       \\
    \hline
    $\chi^2$/d.o.f.    & $374.7/306$              & $731.1/568$                & $733.3/569$
                                                  & $725.0/569$                & $729.4/567$               \\
    \hline
\end{longtable}

\begin{figure}
   \begin{center}
     \FigureFile(120mm,80mm){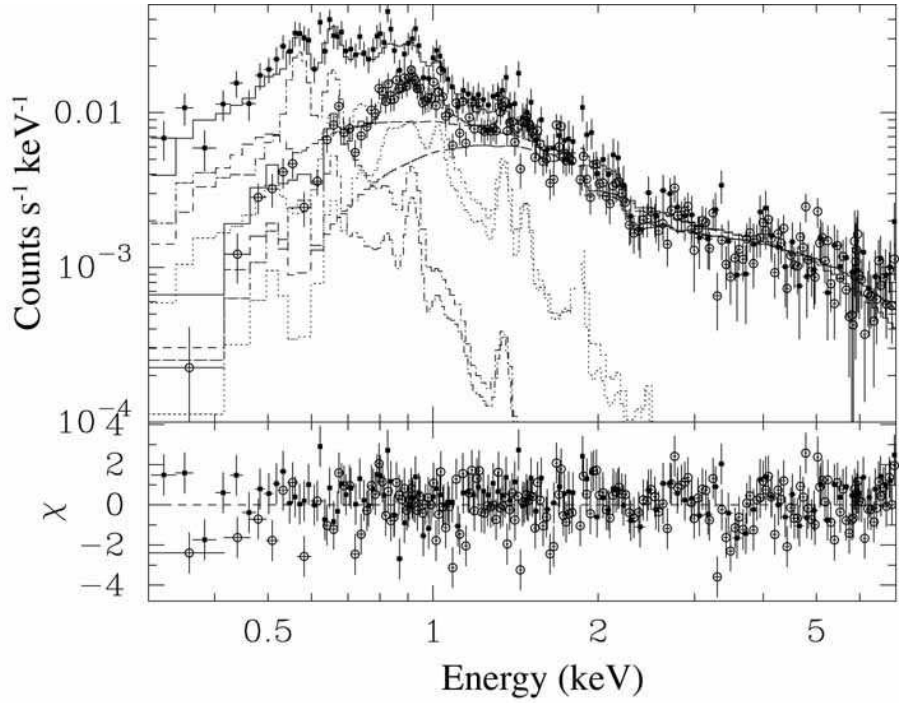}
   \end{center}
   \caption{The XIS spectra of 'BGD-NTE' are shown with the best fit
     model spectra. Black squares and open circles indicate the BI
     (XIS1) spectrum and the sum of the three FI (XIS0,2,3) spectra 
     respectively. The normalization is given for one sensor.
   }
   \label{fig:SuzakuMethod2_BGD_NTE_fit}
\end{figure}

\begin{figure}
   \begin{center}
     \FigureFile(120mm,80mm){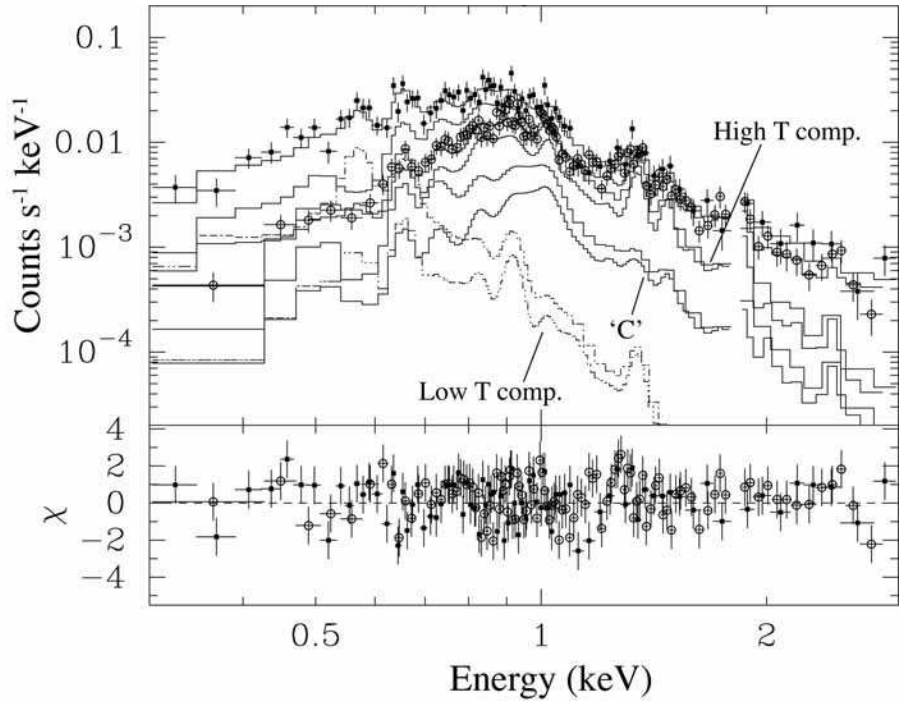}
   \end{center}
   \caption{The XIS spectra of 'CAP-NTE' are shown with the best fit
     model spectra of the two thermal plasma components of the Cap.
     Black squares and open circles indicate the BI (XIS1) spectrum and
     the combined spectrum of the three FI (XIS0,2,3), respectively.
     The normalization is given for one sensor.
   }
   \label{fig:SuzakuMethod2_CAP_NTE_fit}
\end{figure}

\subsection{Uncertainty due to the Plasma Codes}
The spectra obtained with XIS and EPIC show that the emission originates in 
two thermal plasma components. We first discuss the reliability
of the results. We first tried the two-temperature vAPEC model (2vAPEC)
for the fitting of the spectra instead of vMEKAL
\citep{SmithApJL01_APEC}. The reduced $\chi^2$ are almost the same between
the two models. The temperatures, absolute metal abundances and their
ratios agree with each other within the
statistical errors although the absolute metal abundances obtained
with the 2vAPEC model are somewhat lower than those with the 2vMEKAL
model, except for Ne. Thus, there is no obvious reason for adopting one
or other of the two models. Following the previous spectral studies of M82
and the Cap \citep{ReadMN02_M82_RGS, StevensMNL03_M82_EPIC,
OrigliaApJ04_M82_Star_Gas_Abundance}, we consider the  adoption of
vMEKAL as the plasma model in this paper.

\subsection{Uncertainty due to the Contaminant on the OBF of the XIS sensors}
As already mentioned, the thickness of the contaminant on the XIS OBFs 
still has some uncertainty at the time of writing
\citep{Koyama2006XIS}. The uncertainty of the column density of the
contaminant affects the effective area in the soft energy band mainly
below $\sim 1$~keV. In order to check its impact on the
determination of the temperatures and metal abundances, we conducted
spectral fits with 2vMEKAL by changing the thickness of the
contaminant by $\pm 20$\%. The fit gave acceptable results. We
found no change ($<0.1$\%) in the temperature of the high plasma
temperature component. The temperature of the low plasma temperature
component varied by $\pm 5$\%, which is comparable with the statistical
errors. The absolute metal abundances were affected at the level of 
$\pm 9$\% (O),
$\pm 2$\% (Ne), $\pm 2$\% (Mg), $\pm 1$\% (Si) and $\pm 2$\% (Fe), all
of which are significantly smaller than the statistical errors. Thus,
we conclude that the uncertainties due to the contaminant is generally
negligible compared with the statistical errors. We note that nothing
affects the value of the high temperature component
because an X-ray spectrum with a temperature of \KT$\sim 0.6$~keV
is line dominated; hence the temperature determination is driven 
by the line energies, particularly the Fe-L centroid.

\subsection{Uncertainty due to the Number of Thermal Components}
The spectral analyses above show that two thermal plasma components
are required to represent the observed spectra of XIS and EPIC. This
would imply that the plasma of the Cap consists essentially of
multi-components. Thus, one would suspect that the results from the
spectral fits would be significantly affected by increasing the
number of thermal components.

\begin{figure}
   \begin{center}
     \FigureFile(120mm,80mm){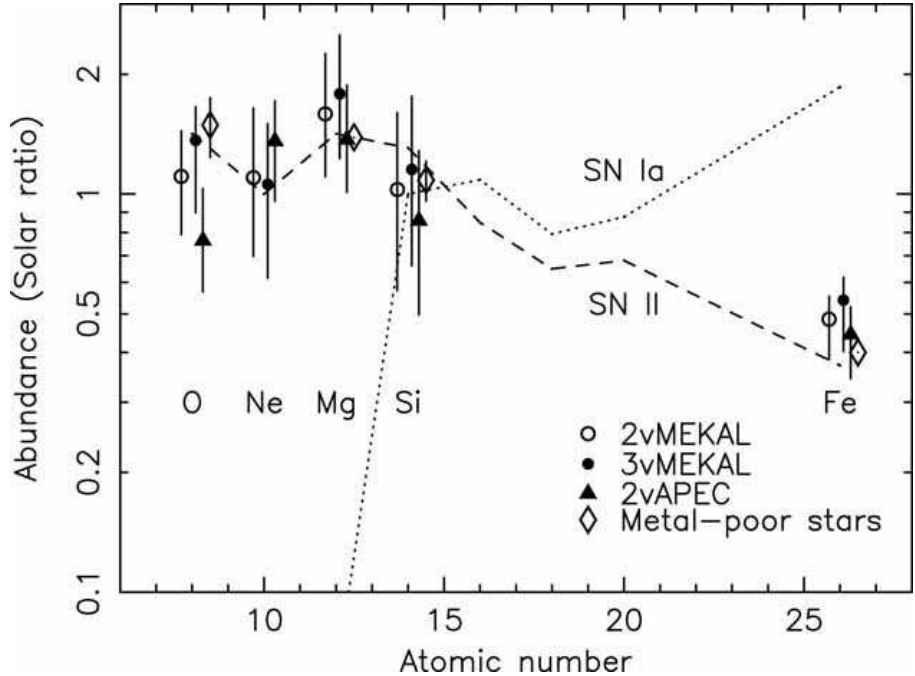}
   \end{center}
   \caption{Metal Abundances as a function of atomic number derived
     from the spectral fits.  The dotted and dashed lines show the
     abundance ratios among metals synthesized by type-Ia and II
     supernovae, respectively. The average metal abundance ratios of
     metal-poor stars are also given, with the normalization such that the
     Fe abundance is $0.4$ solar.
   }
   \label{fig:MetalAbundances}
\end{figure}

In order to check this, we made a fit with a model consisting of three
plasma components (3vMEKAL) and found resulting temperatures of
\KT$=0.13$, $0.44$ and $0.71$~keV (table~\ref{tab:SimultFit}). 
The reduction of $\chi^2$ is only $3.9$ against the number of the 
additional free model parameters of 2, meaning the introduction of 
the third component is not significant from the statistical point of view. 
The temperature and emission integral of the component with \KT$=0.13$~keV
are consistent with those of the low temperature component of the
two-temperature model within the statistical errors. In the
three-component model, the sum of the emission measures and the
weighted mean temperature with the emission integrals of the medium
and high temperature components are almost the same as those of the
high temperature component of the two-component model. The absolute
metal abundances obtained with the three-component model are slightly
higher than those of the two-component model 
(figure~\ref{fig:MetalAbundances}). Nevertheless, the differences
between the two models are within the statistical errors.

In spite of the fact that the model consisting of a single thermal 
component plus a narrow gaussian
line was physically rejected, its resulting metal abundances are also
consistent with those with the 2vMEKAL model within the statistical
errors (table~\ref{tab:SimultFit}). The derived temperature is
consistent with that of the main component obtained in the 
two-temperature model. Thus, we can safely conclude that the results on
the metal abundances and temperatures are robust even if the plasma of
the Cap consists of single or multi components.

\subsection{Non-Equilibrium Ionization}

In the section~\ref{SubSect:SimpleBGDSubt}, we argued that O~VII K
emission line is not explained by a single thermal plasma model with
the temperature of \KT$\sim 0.6$~keV well describing the overall
structure of the observed spectra. It is because the only a small
fraction of O stays in the ionization state of O~VII for the
temperature of \KT$\sim 0.6$~keV in the equilibrium ionization. Thus,
the K emission line from the relatively low ionization state of O~VII
might suggest the plasma is in a non-equilibrium ionization state.

In order to investigate it, we made a spectral analysis with a
non-equilibrium ionization model with the simultaneous fitting 
to the Cap and the blank sky (method 2). We adopted the version 
2.0 of the vNEI model incorporated in XSPEC version 11.3 
\citep{BorkowskiApJ94_Kepler, BorkowskiApJ01_NEI_SedovSNR, 
  HamiltonApJS83_NEI_SNR, LiedahlApJ95_FeL} 
and fixed the whole absorption column density to be 
\NH$=4.0\times 10^{20}$~\UNITNH. The metal abundances of O, Ne, Mg,
Si and Fe were left free as in the perevious sections. 

\begin{figure}
   \begin{center}
     \FigureFile(120mm,120mm){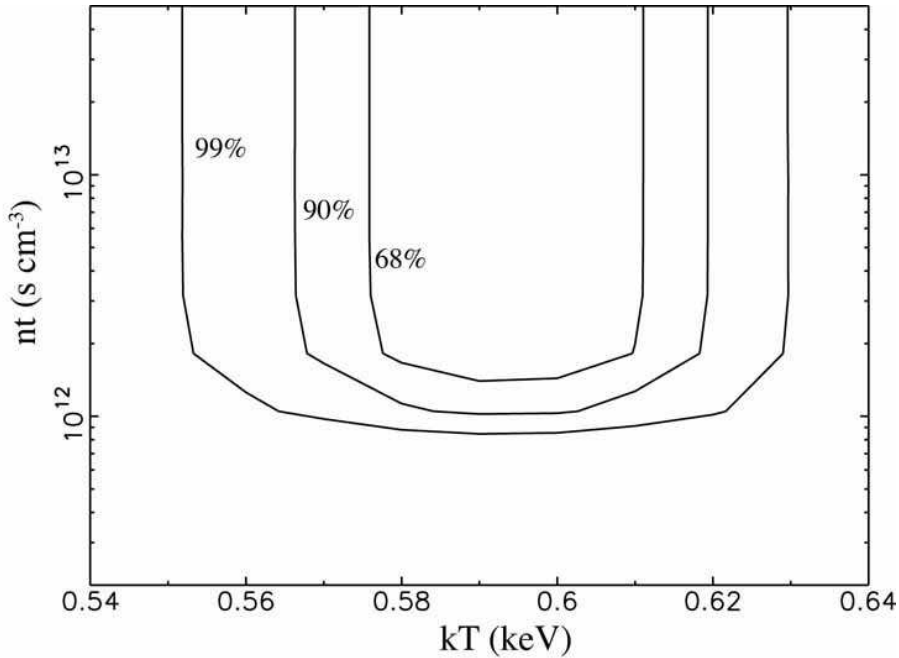}
   \end{center}
   \caption{The confidence contour map of the electron temperature ($kT$)
     and the ionization timescale ($nt$) obtained from the fitting with 1vNEI. 
     The ionization time scale is defined in an unit of ${\rm s\ cm^{-3}}$. 
     The contours are given in confidence levels of 68\%, 90\% and 99\%.
  }
   \label{fig:neiSX1ondo_Steppar}
\end{figure}

Figure~\ref{fig:neiSX1ondo_Steppar} shows the confidence contour map
for the ionization timescale ($nt$) as a function of the electron
temperature (\KT) in the 1vNEI model. We obtained the lower limit of
$8.5\times 10^{11}~{\rm s\ cm^{-3}}$ on the ionization timescale at
the 99\% confidence level, which suggests the observed spectra is
essentially consistent with an equilibrium ionization state. The
significant positive residual at 0.57~keV consistent with the O~VII K
emission line was detected again. Thus, it leads us to conclude that
the model of a non-equilibrium ionization plasma dose not explain 
the O~VII K emission line and the other spectral structures at the same time.

\section{Discussion} 
\subsection{Metal Abundance Ratios}
Using the XIS and EPIC data, we demonstrated that the X-ray plasma of
the Cap region consists of two thermal components.
The high temperature component has \KT$=0.63\pm
0.02$~keV, which agrees with the previously reported ones of
\KT$=0.80\pm 0.17$~keV and \KT$=0.65^{+0.04}_{-0.03}$~keV
\citep{Lehnert1999_ROSAT_cap, StevensMNL03_M82_EPIC}. The emission
integral of this component is also consistent with other observations. The
existence of the O~VII emission line and the low temperature component
with \KT$=0.2$~keV is a new finding of this paper. It would reflect
the nature of the plasma in the Cap region. However, we do not go into
details in this paper.

The XIS image shows the existence of  diffuse emission in the
connecting region between M82 and the Cap
(figure~\ref{fig:Suzaku_BI_Image}), as already pointed out by
\citet{StevensMNL03_M82_EPIC}. This may be expected from a large number of
sequential supernovae, mostly type-II as a result of starburst
activity occurring in M82. Thus, one would expect that the X-ray plasma
in the Cap originates from the type-II supernovae in M82, a hypothesis 
which can be tested with the pattern of relative ratios of the metal 
abundances.

\begin{figure}
   \begin{center}
     \FigureFile(150mm,120mm){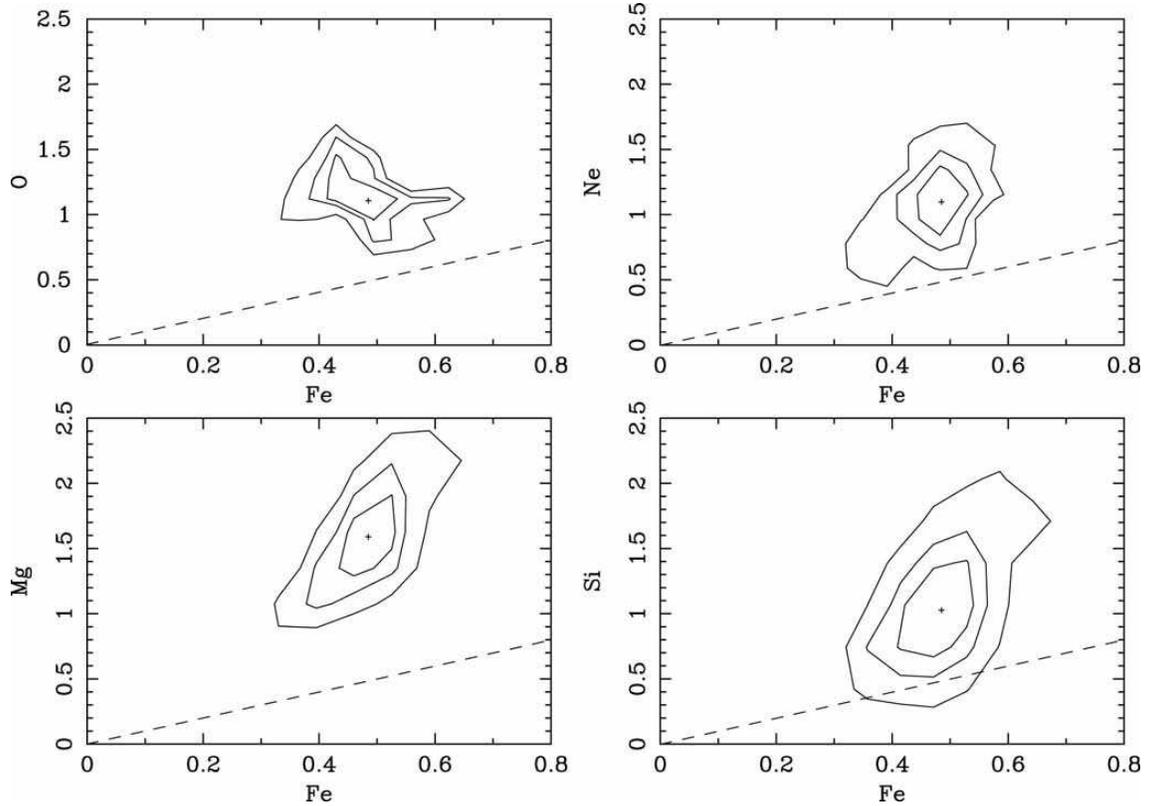}
   \end{center}
   \caption{The confidence contours for the O, Ne, Mg and Si abundances
     relative to the Fe abundance, obtained from the fits using 
     the 2vMEKAL model. The contours are given at confidence levels of
     68\%, 90\% and 99\%. The dashed lines show the solar abundance
     ratio between the metals.
   }
   \label{fig:Fe_O_Mg_Ne_Si_ConfCont}
\end{figure}

Thanks to the large effective areas and good energy resolution of the XIS
and EPIC in the soft X-ray band, we successfully derived the
metal abundances of O, Ne, Mg, Si and Fe in the X-ray plasma of the Cap
for the first time. The abundances of O, Ne, Mg and Si are $1\sim 2$
solar, while that of Fe is about 0.5 solar, which is generally
consistent with the idea of the plasma originating from type-II
supernovae. The emissivities of the lines depend on the plasma
temperature, determined by the shape of the Fe L line complex,
suggesting that the abundances obtained might be strongly coupled with
the Fe abundance. Therefore, we have checked the reliability of the results.
Figure~\ref{fig:Fe_O_Mg_Ne_Si_ConfCont} shows the confidence 
contours for the measured
abundances of  O, Ne, Mg and Si against Fe. This shows
that the metal abundance ratios of O, Ne and Mg to Fe are higher
than unity in units of the solar abundance, at a confidence level of
99\%. An abundance ratio of unity between Si and Fe is permitted at
a confidence level of 99\%, but still rejected at the level of 90\%.
Thus, it is safely concluded that the Fe abundance is significantly
lower than the others.

We have compared the metal abundance ratios of the Cap
with those synthesized by type-II and type-Ia supernovae
(figure~\ref{fig:MetalAbundances}). We adopted the W7 model in
\citet{Nomoto1984_SN1a_Metal} and \citet{Thielemann1986_SN1a_Metal}
for the metal masses synthesized by type-Ia supernovae. The model plot for
type-Ia supernovae normalized with the Si abundance at the solar value shows
clear disagreement with our observational result. This result is
easily expected since O, Ne of Mg are not synthesized by type-Ia
supernovae, while the metal abundances actually observed are much
higher than that of Fe.

For type-II supernovae, we used the results of
$m_u=50$~\UNITSOLARMASS\ by \citet{Tsujimoto1995_SN2_Metal}, in which
the averages of the synthesized masses are 1.8, 0.23, 0.12, 0.12,
0.041, 0.0080 and 0.0091~\UNITSOLARMASS\ for O, Ne, Mg, Si, S, Ar and
Fe, respectively. We plot the metal abundance ratios synthesized by
type-II supernovae, normalized with the Ne abundance at the solar value, in
figure~\ref{fig:MetalAbundances}. Comparison with our results
shows that metal synthesis by type-II supernovae agrees well with
the observed metal abundance ratios in the Cap.

Another good (experimental) sample for metal synthesis by type-II
supernovae comes from observations of metal-poor stars;
these observations are free from the ambiguities resulting 
from the different supernova models. The
chemical composition of these old stars reflects that of the gas from
which they formed in the early universe, where type-II supernovae
dominated type-Ia. Table~6 in \citet{ClementiniMN99_Ab_MetalPoorStar}
shows the average metal abundance ratios in metal-poor stars obtained
with Hipparcos and ground-based observations. After the adjustment of
the definition of the solar abundance, the metal abundance ratios
relative to Fe in metal-poor stars are $3.1\sim 4.4$, $3.5$ and
$2.4\sim 3.0$ for O, Mg and Si, respectively. 
Figure~\ref{fig:MetalAbundances} shows that
the metal abundance ratios of the metal-poor stars and the Cap agree
well with each other. 
Note that the absolute metal abundances of the metal-poor stars 
in the figure are meaningless since they are normalized 
so that the abundance of Fe is $0.4$ solar. 

Thus, the results of the comparison with models of type-II
supernovae and metal-poor stars are consistent with the idea that
the X-ray plasma in the Cap originates from type-II supernovae. 
On the other hand, type-Ia supernovae are not a major contributor to
the metals in the Cap. Thus, this naturally leads to the suggestion that
most of the metals ejected by M82 are synthesized by type-II
supernovae occurring through its starburst activity.

\subsection{Comparison with the Mass of the Dust in the Cap}
\citet{Hoopes2005_GALEX_M82_N253} detected UV emission at the Cap with
GALEX. They suggest that the most likely UV emission mechanism is
scattering of stellar continuum from the starburst in M82 by dust in
the Cap. They also argue that the dust may have been pushed out of M82
by the starburst, or stripped from either M82 or M81 by the tidal
interaction between the two galaxies. The former picture is supported
by the SCUBA observation showing the outflowing dust from M82
\citep{AltonAA99_SBG_DustOutFlow}. The Spitzer result by
\citet{EngelbrachtApJL06_M82_MIR} preferentially supports 
the latter assumption.

Taking the composition of dust grains into account, Si and/or Fe with
significant masses could exist in the form of dust in the Cap. Since the
mass of the dust is not reported by \citet{Hoopes2005_GALEX_M82_N253},
it is impossible to make direct comparison between the metal masses in
the X-ray plasmas and those in the dust at the moment. An upper
limit on the \HI\ column density of $2.7\times
10^{19}$~\UNITNH\ is given by \citet{Yun1993_M82_HI} and
\citet{Yun1994_M81group_HI}. Adopting a volume of $3.7\times 3.7\times
0.9{\rm kpc}^3$ for the emission region of the Cap following
\citet{Lehnert1999_ROSAT_cap}, the upper limit on the mass of the
\HI\ cloud is $\sim 7\times 10^5$~\UNITSOLARMASS. Assuming a
gas-to-dust ratio of 100, the upper limit on the dust mass is $\sim
7\times 10^3$~\UNITSOLARMASS.

Dust grains are sputtered in an X-ray plasma with a time scale of $t_{\rm
   sp} \sim 10^8\left(a/0.1{\rm \mu m}\right)\left(n/10^{-3}
~{\rm cm^{-3}}\right)^{-1}{\rm yr}$ \citep{YamadaPASJ05_Dust_ICM}, where
$a$ and $n$ are the size of the dust grains and the density of the
plasma, respectively, where $f_{\rm x}$ is the filling factor. 
Estimating 
$n\sim 5\times 10^{-3}~{\rm cm^{-3}}\cdot f_{\rm x}^{-0.5}$,
the sputtering time scale is 
$t_{\rm sp}\sim 2\times 10^7{\rm yr} \cdot f_{\rm x}^{0.5}$.
This is comparable with the travel time scale of $\sim 1.5\times 
10^7{\rm yr}$ from M82 to the Cap at the shock speed of 
$\sim 740~{\rm km\ s^{-1}}$, where this speed is required 
to produce the plasma with \KT$\sim 0.63$~keV,
following the estimation in \citet{Lehnert1999_ROSAT_cap}. This
suggests that some fraction of the dust grains have been possibly
sputtered and mixed into the X-ray plasmas.

The masses of Si and Fe in the plasma phase obtained with our
observation are $1.4\times 10^3\cdot f_{\rm x}^{0.5}$~\UNITSOLARMASS\ 
and $1.8\times 10^3\cdot f_{\rm x}^{0.5}$~\UNITSOLARMASS, 
respectively, where $f_{\rm x}$ is the filling factor. 
The sum of these masses is comparable with the upper
limit on the mass of the dust. Thus, it would seem possible that the
dust could contribute a significant part of the metals in the X-ray
plasmas. In other words, dust depletion due to sputtering could
have an impact on the measured abundances from our X-ray results. 
Quantitative observation of
the dust is crucial to understanding the ejecta from starburst activity
and its fate. Further analyses and observations of the dust in the UV
and submillimeter bands are very important.

\subsection{Line Emission through The Charge-Exchange Process}
\citet{LallementAA04_ChargeExchange} has pointed out that the
charge-exchange process could be important in X-ray line emission from
the Cap, where the ionized superwind from M82 can be assumed to
collide with cool ambient gas located at the Cap, as suggested by
\citet{Lehnert1999_ROSAT_cap}. The charge-exchange process contributes
only emission lines to the X-ray spectrum. However, we have detected
continuum emission in the observed X-ray spectra of the Cap. This
observational result suggests that charge-exchange is not the major 
process in the Cap, although a minor contribution cannot be denied.

\begin{figure}
   \begin{center}
     \FigureFile(120mm,120mm){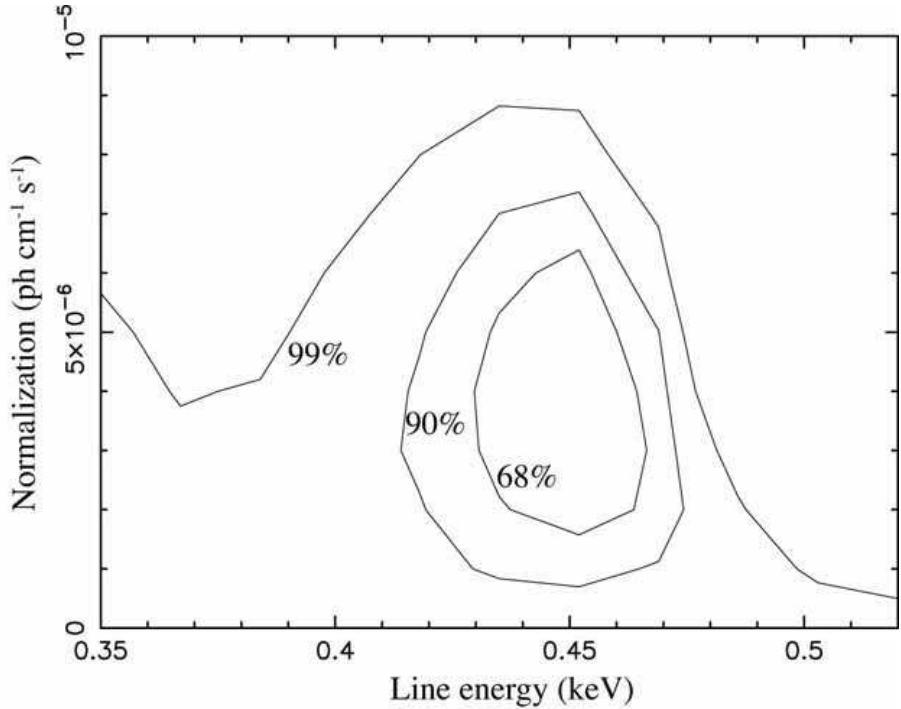}
   \end{center}
   \caption{The confidence contour map of the C~VI emission line
     energy at 0.459~keV and its normalization obtained from the fitting with 
     2vMEKAL $+$ 1GAUS. 
     The normalization of the line is defined as an unabsorbed photon flux,
     where the normalization unit is ${\rm ph\ cm^{-2}\ s^{-1}}$.
     The contours are given in confidence levels of 68\%, 90\% and
     99\%.
   }
   \label{fig:CVIconfCont}
\end{figure}

According to the charge-exchange emission model for comets by
\citet{KrasnopolskySSRv04_Comet_Xray_EUV}, the $n=4$ to $1$ transition
line is enhanced for C~VI at 0.459~keV. During the charge-exchange
process, the electron from the donor neutral atom is first trapped at
high levels, typically $n=4$ for carbon. The Suzaku and XMM-Newton
spectra given in figures~\ref{fig:SuzakuSimpleBGDsubtSpec},
\ref{fig:SuzakuMethod2_CAP_NTE_fit} and \ref{fig:XMMSpec} show a hint
of an emission line around  0.46~keV. Thus, we added a narrow
gaussian line around 0.459~keV to the 2vMEKAL model and examined
its confidence. Figure~\ref{fig:CVIconfCont} shows the confidence
contour map for the narrow gaussian line. A local minimum is found
at an energy consistent with 0.459~keV, which suggests marginal
detection of the C~VI emission line although it is not statistically
significant at the 99\% confidence level.

We have estimated possible contributions to emission lines, especially O, 
due to the charge-exchange process. Following the model of
\citet{Lehnert1999_ROSAT_cap}, we again assume a box with dimensions of
$3.7\times 3.7\times 0.9{\rm kpc}^3$ for the Cap. An upper limit
of $2\times 10^{-3}{\rm cm}^{-3}$ has been obtained for the density of
the \HI\ cloud at the Cap \citep{Yun1993_M82_HI, Yun1994_M81group_HI}.
Thus, the upper limit on the column density of the \HI\ cloud along
the direction of the superwind from M82 is estimated to be $8\times
10^{18}\ {\rm cm}^{-2}$. Since the cross-section of the
charge-exchange between an O ion and a H atom is 
$\sim 10^{-15}-10^{-14}\ {\rm cm}^2$ 
\citep{WegmannPSS98_CometXray_ChargeExchange},
it is expected that an O ion in the superwind  would encounter 
a \HI\ cloud thick enough for the charge-exchange process. 

In this paper, we will not discuss the details of the process 
by which the superwind
and O ions collide with the \HI\ cloud. Nevertheless, we can
estimate an upper limit on the photon flux of O emission lines
assuming an extreme case in which all O ions suffer the charge-exchange
process with \HI\ and emit O K emission lines. Assuming a density of
$\sim 1\times 10^{-3}\ {\rm cm^{-3}}$, the velocity of the superwind
of $\sim 740~{\rm km\ s^{-1}}$ and solar abundance of O in the plasma,
the upper limit on the photon flux is estimated to be $5\times
10^{-6}$~\UNITPFLUX. The observed unabsorbed photon flux of O~VII is
$\sim 6\times 10^{-6}$~\UNITPFLUX (tables~\ref{tab:SimultFit}). We
note that the same logic gives an upper limit of $2\times
10^{-6}$~\UNITPFLUX\ on the photon flux of C due to the charge-exchange
process, and this is consistent with the upper limit on the
one possible C~VI emission line at 0.459~keV. Thus, the upper limit on the
photon flux of the O K emission lines due to charge-exchange is at the
same order of the one actually observed. This suggests that the
charge-exchange process may make a significant contribution to the 
observed O K emission lines.

We thus expect that the charge-exchange process could be one of the key 
aspects of the understanding of the physical processes and the 
measurement of the metal abundances in the Cap, 
and hence the ejecta from starburst galaxies and its cosmic evolution. 
More extensive studies of the possible charge-exchange are 
definitely necessary. This could not be achieved with the limited 
spectral resolution of X-ray CCDs. The realization of the
non-dispersion type of spectrometer with ultra high spectral
resolution, i.e. the $\mu$-calorimeters onboard the NeXT and DIOS
missions is essential \citep{KuniedaSPIE06_NeXT, OhashiSPIE06_DIOS}.

\section{Summary} 
\noindent 1. The Suzaku XIS clearly detected the diffuse X-ray emission
from the Cap region located $\sim 11'$ to the north of the M82
galaxy. Additional diffuse X-ray emission is also detected in the
connecting region between the Cap and the M82 galaxy.\\

\noindent 2. The X-ray spectrum of the Cap is well described by a model
with two thermal plasma components. The fit using the single-temperature 
model plus a narrow gaussian line at 0.57~keV is physically rejected 
because it resulted in a lower absorption column than that of our galaxy. 
Non-equilibrium ionization is not required to fit the spectra. \\

\noindent 3. The metal abundance ratios of the Cap agree well with
those of metal-poor stars and the model predictions of metals
synthesized by type-II supernovae, but are not consistent with the models of
type-Ia supernovae. This result supports the idea that the origin of
the metals in the Cap is type-II supernovae explosions occurring in the
starbursts in the M82 galaxy.\\

\noindent 4. A significant contribution to the metals in the X-ray plasma in
the Cap is possible from the sputtered grain dusts. This depends on the
total mass of the dust.\\

\noindent 5. An emission line consistent with the C~VI transition from
$n=4$ to $1$ at 0.459~keV is marginally detected.  Although this is
not statistically significant, it supports possible charge-exchange
processes in the Cap. The charge-exchange process may also make a 
significant contribution to the observed O K emission lines.

\section*{Acknowledgement}
We are grateful to all the members of the Suzaku hardware and software
teams and the science working group. YH are supported by JSPS Research
Fellowship for Young Scientists. REG acknowledges the support of NASA
grant NNG05GR02G. This work is based by the Grant-in-Aid for the 21st
Century COE ``Center for Diversity and Universality in Physics'' from
Ministry of Education, Culture, Sports, Science and Technology (MEXT)
of Japan, and is supported on a Grant-in-aid (Fiscal Year 2002-2006)
for one of Priority Research Areas in Japan; ``New Development in
Black Hole Astronomy''. This work is partially supported by
Grants-in-Aid of Grants-in-Aid of Ministry of Education, Culture,
Sports and Technology (MEXT) of Japan No. 15340088.


\end{document}